\documentclass[aps,prb,reprint]{revtex4-1}

\usepackage{amsmath}
\usepackage{amsfonts}
\usepackage{subfig}
\usepackage{graphicx}
\usepackage{url}
\usepackage{multirow}
\usepackage{cleveref}

\usepackage{dcolumn}
\usepackage{bm}

\begin{document}

\title{Electronic structure tuning via surface modification in semimetallic nanowires}

\author{Alfonso Sanchez-Soares}
\affiliation{Tyndall National Institute, University College Cork, Dyke Parade, Cork, T12 R5CP, Ireland}
\author{Conor O'Donnell}
\affiliation{Tyndall National Institute, University College Cork, Dyke Parade, Cork, T12 R5CP, Ireland}
\author{James C.~Greer}%
 \email{jim.greer@tyndall.ie}
\affiliation{Tyndall National Institute, University College Cork, Dyke Parade, Cork, T12 R5CP, Ireland}

\begin{abstract} 
Electronic structure properties of nanowires (NW) with diameters of 1.5 nm and 3 nm based on semimetallic $\alpha$-Sn are investigated by employing density functional theory and perturbative $GW$ methods. We explore the dependence of electron affinity, band structure and band gap values with crystallographic orientation, NW cross-sectional size and surface passivants of varying electronegativity. We consider four chemical terminations in our study: methyl ($\rm{CH_3}$), hydrogen ($\rm{H}$), hydroxyl ($\rm{OH}$), and fluorine ($\rm{F}$). Results suggest a high degree of elasticity of Sn-Sn bonds within the SnNWs' cores with no significant structural variations for nanowires with different surface passivants. Direct band gaps at Brillouin zone centers are found for most studied structures with quasi-particle corrected band gap magnitudes ranging from 0.25 eV to 3.54 eV in 1.5 nm diameter structures indicating an exceptional range of properties for semimetal NWs below the semimetal-to-semiconductor transition. Band gap variations induced by changes in surface passivants indicate the possibility of realizing semimetal-semiconductor interfaces in NWs with constant cross-section and crystallographic orientation allowing the design of novel dopant-free NW-based electronic devices.
\end{abstract}

\maketitle

\section{\label{sec:Intro}Introduction}

The nanopatterning of solids has been shown to induce significant variations in material properties with respect to their bulk counterparts as a result of quantum confinement effects and an increase in the impact of surface phenomena on the system as a whole due to their large surface-to-volume ratios. Tuning of a broad range of properties has been successfully demonstrated and the design of novel nanostructures exploiting chemical and structural modifications have allowed elucidation of numerous applications of such as chemical sensors,\cite{Liao2009,Kim2009} transparent electronics,\cite{Dattoli2007} and tunnel FETs;\cite{Zhao2015,Conzatti2012} making exploitation of size effects a way of enhancing a material's versatility by allowing some degree of tunability of its properties through geometric effects.

The study of semiconductor nanowires (NW) structures has been increasing since \citet{Wagner1964} first proposed the vapor-solid-liquid growth mechanism and demonstrated  submicron-scale wires of semiconducting materials five decades ago. More recently, with the advances in nanofabrication, NWs with diameters as low as 1 nm have been experimentally realized\cite{Ma2003} and their measured band gaps reproduced by $GW$-corrected density functional theory (DFT) calculations.\cite{Read1992,Zhao2004} Analog to band gap \emph{widening} observed in semiconductors, bulk semimetals' electronic properties have been reported to transition to those of a semiconductor with the emergence of a band gap for structures with critical dimensions on the order of 10 nm as a result of quantum confinement with predicted band gap values well above 1 eV in $\alpha$-Sn NWs with diameters of approximately 1 nm.\cite{Ansari2012}

Further control of NW band gaps has been reported in silicon NWs by passivating surface states with chemical groups of varying electronegativities reported to induce band gap variations of the same order of magnitude as quantum confinement effects. \citet{Leu2006} report variations of the order of electronvolts in SiNW band gaps when passivated with halogens with respect to hydrogen-passivated structures, attributing reductions in band gaps to weakly interacting surface species. \citet{Nolan2007} studied band gap variations on similar systems passivated with $\rm{NH_2}$ and $\rm{OH}$ and correlated band gap variations of up to 1 eV to orbital hybridization between passivants and NW core; while \citet{Zhuo2013} correlated such band gap reductions in SiNWs to charge redistributions and corresponding electrostatic effects induced by surface chemistry.

In the following, we explore the effects of surface passivants with varying electronegativity on NWs based on semimetallic $\alpha$-Sn. Although not stable in bulk at room temperature\cite{Paul1961}, studies show Sn's $\alpha$ phase can be stabilized in thin film structures~\cite{Farrow1981,Hochst1983,John1989,Ueda1991} while $\alpha$-Sn nanocrystals have been the focus of recent studies for applications in lithium ion batteries.\cite{Im2013,Kufner2013,Oehl2015} Due to its simple crystal structure we study $\alpha$-Sn as a test system for exploring the range of properties that can be obtained from the competing effects of quantum confinement and chemical passivation in sub-5-nm structures based on bulk semimetals. We study the evolution with NW size and crystallographic orientation by considering structures with diameters of approximately 1.5 nm and 3 nm grown along low index crystallographic orientations. As NW structures have been reported to exhibit electron transport properties which makes them particularly suitable for applications in electronics\cite{Appenzeller2008} and in particular $\alpha$-Sn NWs have been previously proposed for the design of novel nanoelectronic devices,\cite{Ansari2012} we take a particular interest in the effects on energy band gaps --a key quantity for such applications-- and improve upon the DFT description by means of a scheme based on the $GW$\cite{Hedin1965} approximation which has been shown to greatly improve the description of band gaps in semiconductors.\cite{Schilfgaarde2006,Gruning2006}

\section{\label{sec:Method}Method}

\begin{figure*}

\includegraphics[scale=1.0]{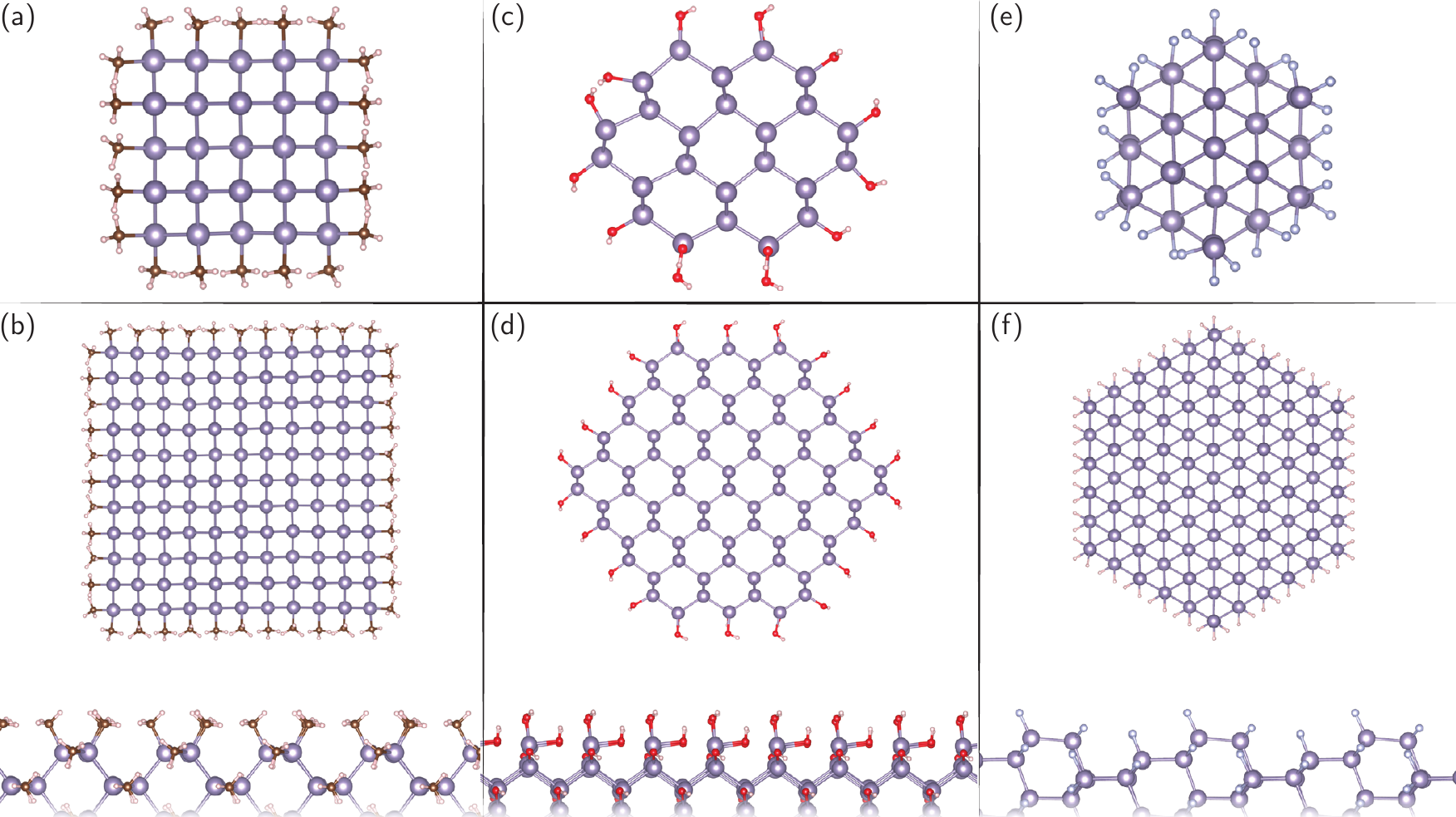}

\caption{Atomistic illustrations of SnNW cross-sectional and side views of optimized NW structures of 1.5 nm (top row) and 3 nm (bottom row) diameters. (a)-(b) $[ 100 ]$ $\rm{SnCH_3}$, (c)-(d) $[ 110 ]$ SnOH, and (e)-(f) $[ 111 ]$ SnF. }\label{fig:struct}

\end{figure*}

Nanowire structures with with cross-sectional areas corresponding to cylindrical structures with diameters of approximately 1.5 nm and 3 nm are modelled from bulk $\alpha$-Sn's diamond structure (space group $\rm{Fd \bar 3m}$, no. 227). NWs oriented along the $[100]$, $[110]$, and $[111]$ crystallographic orientations are considered with their surface bonds passivated by saturating the bonding for the group IV Sn atoms. Four different monovalent passivating groups with varying degrees of electronegativity are chosen for the study of the effects of surface passivation: methyl (-CH$_3$), hydrogen (-H), hydroxyl (-OH), and fluorine (-F).  Due to the large surface-to-volume ratio in these nanostructures a significant proportion of the Sn atoms which constitute the NWs directly bond to the surface passivants, accounting for 40\% to 44\% of the chemical composition of the system as a whole in the 1.5 nm NWs, and approximately 26\% to 27\% in the 3 nm NWs dependent on the crystallographic orientation of a given NW. 
 
Density functional theory (DFT)\cite{Dreizler1990,Soler2002,QW} within the usual Kohn-Sham\cite{Kohn1965} framework is employed to simulate the electronic structure of these nanostructures. We take the supercell approach for computing the properties of systems with reduced dimensionality by which at least 15 \AA ~of vacuum is introduced along directions perpendicular to the NWs' axes in order to eliminate interactions with periodic images normal to the NWs' axes, leading to a simulation model for the NWs as infinitely long and perfectly crystalline quasi-1D systems. The PBE\cite{PERDEW96} formulation of the generalized gradient approximation (GGA) for the exchange and correlation functional is used in conjunction with norm-conserving pseudopotentials\cite{Morrison1993} and double-$\zeta$-polarized basis sets of localized pseudo-atomic orbitals\cite{OZAKI03,OZAKI04}. Brillouin zone integrations are performed over a grid of k-points generated according to the Monkhorst-Pack\cite{Monkhorst1976} scheme with a density of approximately 25 k-points $\times \rm{\AA}$ along a NW axis, whilst real-space quantities are discretized on a grid with corresponding energy cut-offs of 200 Ry (280 Ry) for NWs passivated with hydrogen and methyl (fluorine and hydroxyl). In this approximation the equilibrium cell parameter and bulk modulus of $\alpha$-Sn are predicted to be 6.70 \AA ~ and 38 GPa, respectively. This represents a 3\% deviation from experimentally reported values of 6.49 \AA~ in the equilibrium cell parameter\cite{Thewlis1954,Farrow1981}, and a 30\% underestimation of the bulk modulus with respect to experimentally reported values of 54 GPa\cite{Buchenauer1971,Kamioka1983}. Atomic positions in the NWs are relaxed until forces acting on all atoms are below $10^{-2}$ eV/\AA, and the total energy of the cell is minimized with respect to its length along the NW axis. \Cref{fig:struct} shows cross-sectional and side views of each NW orientation and size with varying surface terminations.

Given the well-known shortcomings of standard DFT in predicting electronic band gaps\cite{Perdew1982,Perdew1983,Gruning2006} -- a key parameter for semiconductors with applications in electronics -- we improve our description through a perturbative many-body approach. We perform first-order perturbative $GW$\cite{Hedin1965,Marini2009,Aryasetiawan1998} quasi-particle corrections for 1.5 nm NWs to gain insight into the order of magnitude of energy band gaps induced as a result of quantum confinement. Electronic structure calculations based on plane-waves\cite{Gianozzi2009} and norm-conserving pseudopotentials\cite{Troullier1991} serve as starting point for computing first-order corrections within the QP approximation\cite{Hybertsen1986} as

\begin{equation}
E_{nk}^{QP} = E_{nk}^{KS}+Z_{nk} [\Sigma_{nk}(E_{nk}^{KS})-V_{nk}^{XC}],
\end{equation}

where $E_{nk}^{KS}$ are the Kohn-Sham eigenvalues computed from the Kohn-Sham procedure, $\Sigma_{nk}$ represents the self-energy which includes exchange and correlation effects, calculated as the convolution of one-electron Green's functions and the dynamically screened Coulomb interaction $W$ within the plasmon-pole approximation\cite{Aryasetiawan1998}; the DFT $V_{nk}^{XC}$ is replaced by the $GW$ correction and the renormalization factor --which accounts for the fact that $\Sigma_{nk}$ is evaluated at Kohn-Sham eigenvalues rather than at $E_{nk}^{QP}$-- is given by

\begin{equation}
Z_{nk} = \left[ 1 - \frac{d \Sigma_{nk}(\omega)}{d \omega} \right]^{-1}.
\end{equation}

Plane-wave calculations for 1.5 nm structures have been performed with kinetic energy cutoffs of 75, 90, 110, and 140 Ry for wavefunction expansions in structures passivated with hydrogen, methyl, fluorine and hydroxyl, respectively. Structural and electronic properties found in simulations performed with localized orbitals were reproduced with calculations relying on plane-wave basis sets. Given the reduced dimensionality of the systems under study, a technique for cutting off long range Coulomb interactions has been employed when computing $GW$  quasi-particle corrections.\cite{Rozzi2006}

\section{\label{sec:Results}Results and Discussion}
\subsection{Nanowire structure}

\begin{table}
\caption{\label{tab:Cell_Parameter}Optimized cell parameter in \AA\ along each NW's periodic direction and characteristic cross sectional length of Sn cores $\phi_{Sn}$. Bulk column indicates cell parameter along corresponding crystallographic directions in bulk $\alpha$-Sn.}
\begin{ruledtabular}
\begin{tabular}{l *{5}{c}}
& Bulk & \multicolumn{2}{c}{1.5 nm} & \multicolumn{2}{c}{3 nm}\\
& $\rm{a_0}$ (\AA) & $\rm{a_0}$ (\AA) & $\phi_{Sn}$ (nm) & $\rm{a_0}$ (\AA) &$\phi_{Sn}$ (nm) \\
\hline \\
$[ 100 ]$ &  &  & \\
\hspace{15pt} $\rm{CH_{3}}$ &\multirow{4}{*}{6.70} & 7.00 & 1.35 & 6.87 & 2.95  \\
\hspace{15pt} $\rm{H}$ & &6.61 & 1.37 & 6.77 & 2.97 \\
\hspace{15pt} $\rm{OH}$ & & 6.11 & 1.43 & 6.40 & 3.04 \\
\hspace{15pt} $\rm{F}$ & & 6.63 & 1.40 & 6.72 & 3.00 \\
$[ 110 ]$ &  &  & \\
\hspace{15pt} $\rm{CH_{3}}$ &\multirow{4}{*}{4.74} & 4.82 & 1.45 & 4.79 & 2.93 \\
\hspace{15pt} $\rm{H}$ & & 4.73 & 1.56 &  4.74 & 2.95\\
\hspace{15pt} $\rm{OH}$ & & 4.71 & 1.58  & 4.74 & 2.96\\
\hspace{15pt} $\rm{F}$ & & 4.75 & 1.59 & 4.76 & 2.98 \\

$[ 111 ]$ &  &  & \\
\hspace{15pt} $\rm{CH_{3}}$ &\multirow{4}{*}{11.60} & 11.74 & 1.54 & 11.69 & 3.00 \\
\hspace{15pt} $\rm{H}$ & & 11.52 & 1.50 & 11.62 & 2.98 \\
\hspace{15pt} $\rm{OH}$ & & 11.31 & 1.54 & 11.52 & 3.00 \\
\hspace{15pt} $\rm{F}$ & &12.04 & 1.52 &  11.77 & 3.05 \\

\end{tabular}
\end{ruledtabular}
\end{table}

Optimal structural parameters are reported in \cref{tab:Cell_Parameter} for the combinations of the SnNW's crystallographic orientations and surface terminations studied.
Small deviations in the cell parameter relative to the corresponding lattice spacing in bulk $\alpha$-Sn are found for NWs passivated with atomic terminations and NWs with larger cross sections. In 1.5 nm NWs both hydrogen- and fluorine-passivated structures exhibit lattice spacings along the NW axis within 1\% of the corresponding bulk value, with the exception of the 1.5 nm $[111]$ structures, in which an elongation of 3.8\% is observed for the fluorine-passivated structure. Larger deviations relative to the corresponding bulk cell parameters are found for thinner structures and NWs passivated with molecules: cell parameters up to 4.5\% larger than the corresponding bulk values are found to be induced by the larger size of the methyl molecules, whilst a combination of the high level of malleability of Sn-Sn bonds and the electrostatic attraction induced by the dipole moment of hydroxyl molecules at the surface favors smaller cell lengths. Despite variations in computed relaxed cell lengths encountered for different surface terminations, overall the NW's total energy displays only a remarkably weak dependence on axial strain as compared with similar strain magnitudes in the bulk. 

Surface terminating groups are found to bond to surface Sn atoms without introducing significant distortions to the NW's Sn atomic structure. For surface facets exposed in $[ 100 ]$- and $[ 111 ]$-oriented structures all terminations are found to bond to the topmost surface Sn atoms in positions exterior to the Sn core. For structures oriented along $[ 110 ]$, there are sites in which  hydroxyl and fluorine are found to align to the NW surface defined by the outermost Sn atoms as the structure is allowed to relax to a minimum energy configuration. This results in Sn-X ($\rm{X=OH, F}$) bonds approximately parallel to the NW axis in $[ 110 ]$-oriented structures, as can be seen in \cref{fig:struct}(c)-(d). The small deviations in optimal cell lengths found for $[ 110 ]$ NWs of 1.5 nm and 3 nm passivated with hydroxyl groups are attributed to this structural rearrangement, as the positioning of hydroxyl groups between Sn planes counteracts dipole attraction forces --which favour shorter cell lengths-- responsible for larger deviations in $[ 100 ]$- and $[ 111 ]$- oriented structures.

Due to variations in NW shape across structures with different crystallographic orientations, and in order to provide a unified measure of NW size, we report as cross-sectional dimensions in \cref{tab:Cell_Parameter} the Sn core diameter ($\phi_{Sn}$) corresponding to perfectly cylindrical structures with cross-sectional areas matching that of each NW structure, where areas have been computed by only taking Sn atoms into account as to preserve comparability across surface terminations. Cross-sectional dimensions are found to remain constant across surface terminations for a given crystallographic orientation even for 1.5 nm structures, as variations reported in \cref{tab:Cell_Parameter} for a given orientation and NW size correspond to differences in Sn-Sn bond lengths in directions perpendicular to the NW axis beyond the accuracy limits of our current approximation. We thus conclude that variations in relaxed cell parameters found for different surface terminations are not followed by concomitant variations in Sn-Sn bond lengths along NW cross sections.

The crystallographic orientation of the structures is found to correlate with the magnitude of maximum deviations in optimal cell parameters with respect to bulk values. NWs oriented along $[ 100 ]$ exhibit the largest variations with molecular terminations inducing deviations of up to 4.5\% and 8.8\% in 1.5 nm structures, whilst structures oriented along $[ 110 ]$ have the smallest deviations with a maximum of 1.7\% for passivation with methyl molecules on the 1.5 nm NWs. As the cross-section of the NWs increases, the magnitude of deviations in the relaxed lattice spacings decreases with the 3 nm NWs exhibiting structural parameters closer to those of bulk $\alpha$-Sn. This variation in structural parameters is attributed to variations in the properties of exposed surface facets with different surface terminations across structures oriented along different crystallographic axes. As NW diameter increases and the surface-to-volume ratio decreases, the impact of surface phenomena on the properties of the whole structure decreases resulting in structural parameters rapidly approaching those of bulk $\alpha$-Sn.

\subsection{Electronic structure}

\begin{figure}

\includegraphics[scale=1.0]{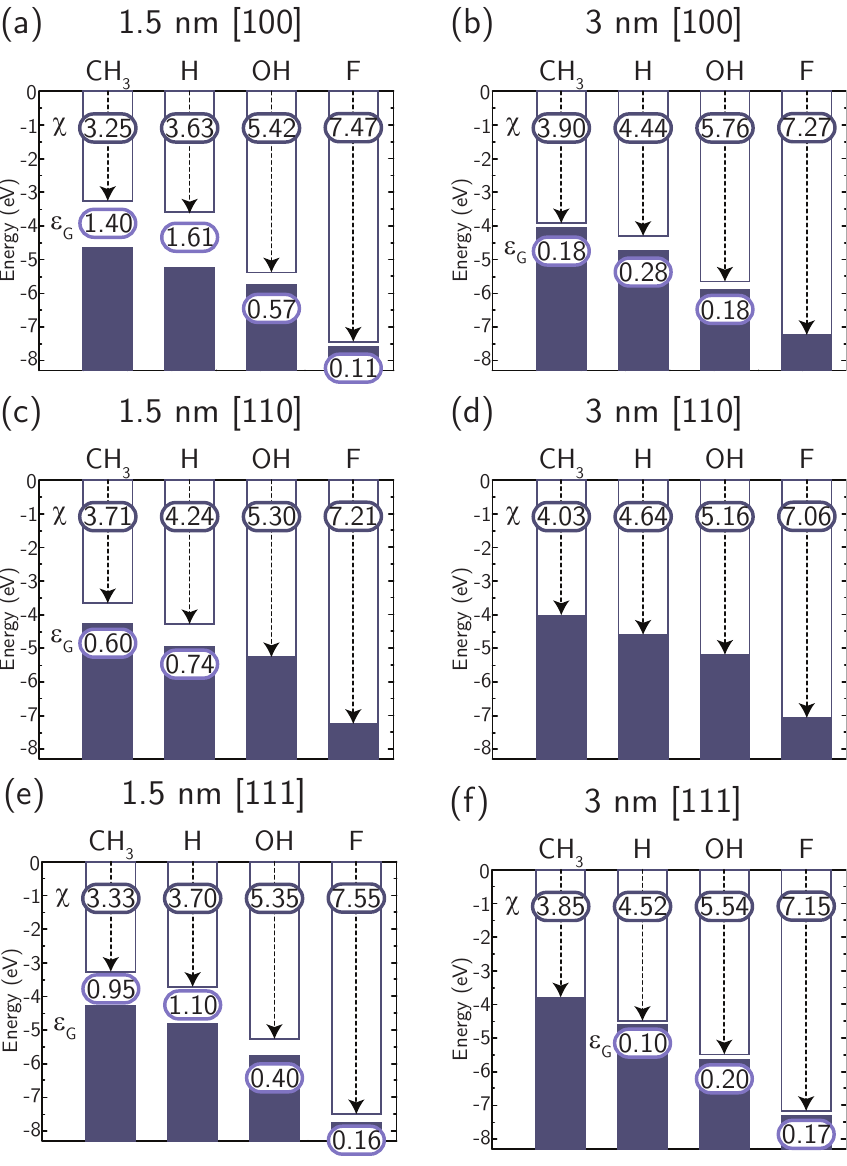}

\caption{Electron affinity and band gap values for 1.5 nm and 3 nm $\alpha$-Sn NWs passivated oriented along (a)-(b) $[ 100]$, (c)-(d) $[ 110 ]$, and (e)-(f) $[ 111 ]$ crystallographic orientations. All reported energies are in eV. Zero of energy is set to vacuum level.}\label{fig:bandgaps}

\end{figure}

\begin{figure}[t]

\includegraphics[scale=1.0]{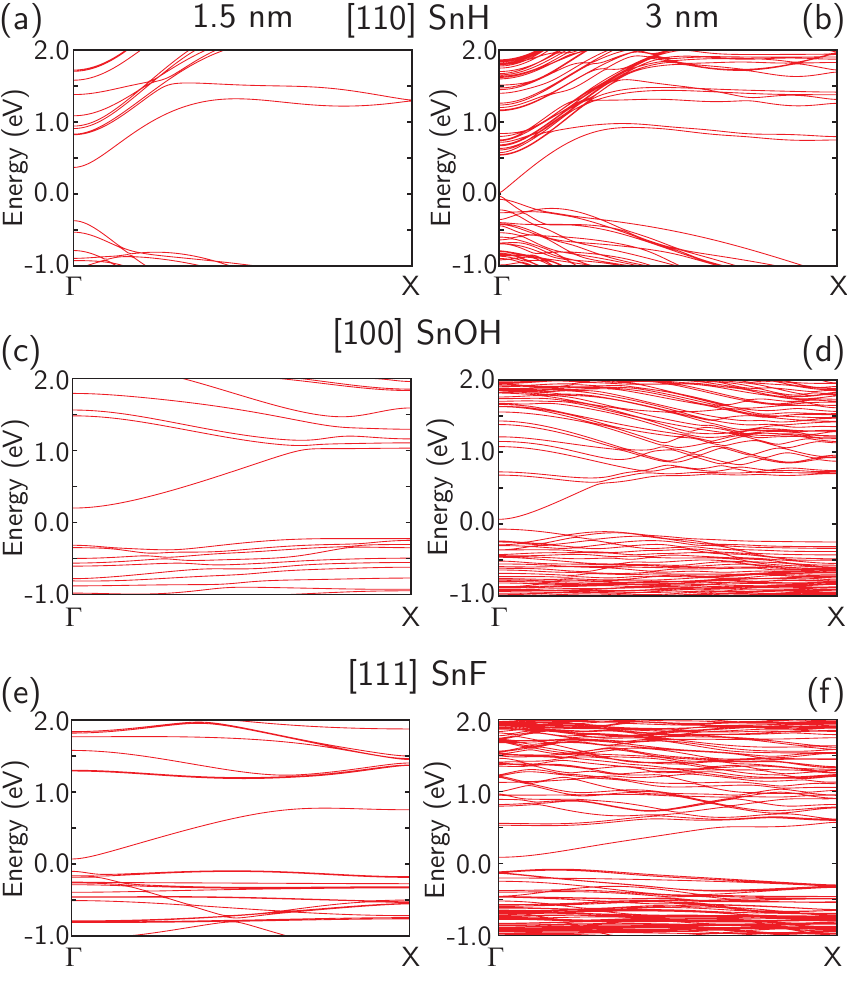}

\caption{Band structures of 1.5 nm and 3 nm $\alpha$-Sn NWs oriented along different orientations and passivated with varying chemical species. (a)-(b) $[ 110 ]$ SnH, (c)-(d) $[ 100 ]$ SnOH, and (e)-(f) $[ 111 ]$ SnF. Zero of energy is set to the Fermi level.}\label{fig:bands}

\end{figure}

The electronic structure of 1.5 nm and 3 nm $\alpha$-Sn NWs is found to strongly depend on crystallographic orientation and surface terminations, with surface effects being more pronounced in thinner structures. \Cref{fig:bandgaps} summarizes computed DFT values for band gaps and electron affinities. 

The calculations reveal that both the band gap and electron affinity can be varied for a given NW cross section and orientation via surface chemistry. Varying the surface termination groups can lead to variations in the band gap energy on the same order of magnitude as induced by quantum confinement. Variations for the SnNW's electron affinities are on the order of electronvolts for surface terminations considered. For confinement dimensions below 5 nm, the potential profile confining electrons within the NWs results in a semimetal-to-semiconductor transition with the emergence of a band gap for hydrogen terminated NWs, which we take as a reference system. The potential acting on the electrons inside the SnNW is found to be strongly related to the terminating species. As the electronegativity of the surface terminating species is increased, a shift relative to the equivalent hydrogen terminated SnNW to the electronic energy levels towards lower values and significant reduction in the energy band gap values occurs. This results in the NW size at which the semimetal-to-semiconductor occurs being dependant on \emph{both} crystallographic orientation --as well as exposed surface facets along cross-sectional directions-- and on the surface termination, with DFT calculations predicting variations between semimetallic behaviour and energy band gaps up to 0.74 eV in 1.5 nm $[ 110 ]$-oriented structures by just varying surface passivants. Results show that for a given NW size and crystallographic orientation, the largest band gaps occur for hydrogen-passivated structures and the smallest band gaps for the highly electronegative case of fluorine, with values for structures passivated with other terminations following trends corresponding to the passivant's electronegativity. An exception to this trend is found for 3 nm $[ 111 ]$ NWs: whilst methyl- and hydrogen-terminated structures approach semimetal behavior and exhibit band gaps below 100 meV, NWs passivated with highly electronegative species are found to exhibit larger band gaps with the fluorine-passivated structure even maintaining the magnitude of its band gap with respect to its 1.5 nm counterpart.

\Cref{fig:bands} shows band structure plots for three representative SnNWs. \Cref{fig:bands}(a)-(b) depicts the electronic structure arising for the case of hydrogen-passivated $[ 110 ]$ NWs in which a direct band gap at $\Gamma$ is found for both 1.5 nm and 3 nm structures, and as anticipated from the quantum confinement effect, the magnitude reducing as the confinement dimensions increase and is reflected in the majority of SnNWs considered in this study. Exceptions are encountered are encountered for the cases shown in \cref{fig:bands}(c)-(f): $[ 100 ]$ NWs passivated with hydroxyl --\cref{fig:bands}(c)-(d)-- are found to exhibit an indirect band gap for 1.5 nm NWs which transitions to a direct gap at $\Gamma$ for the corresponding 3 nm structure, while the opposite situation is encountered in the band structures of hydroxyl- and fluorine-passivated $[ 111 ]$ NWs, which are predicted to exhibit a direct band gap at $\Gamma$ for 1.5 nm and transition to an indirect band gap for the 3 nm SnNW as shown in \cref{fig:bands}(e)-(f) for the case of fluorine. Another feature unique to $[ 111 ]$ SnNWs passivated with highly electronegative terminations is a gap of approximately 0.4 eV predicted to be between the conduction band and bands at higher energies as seen in \cref{fig:bands}(e). This same feature with a gap similar in magnitude has been observed in the band structure of 1.5 nm $[ 111 ]$ SnNWs passivated with hydroxyl, while for less electronegative terminations, no gap is predicted within the corresponding energy range. As can be seen in \cref{fig:bands}(f) this gap reduces with increasing NW size with values of less than 40 meV in the 3 nm structures.

\begin{figure}

\includegraphics[scale=1.0]{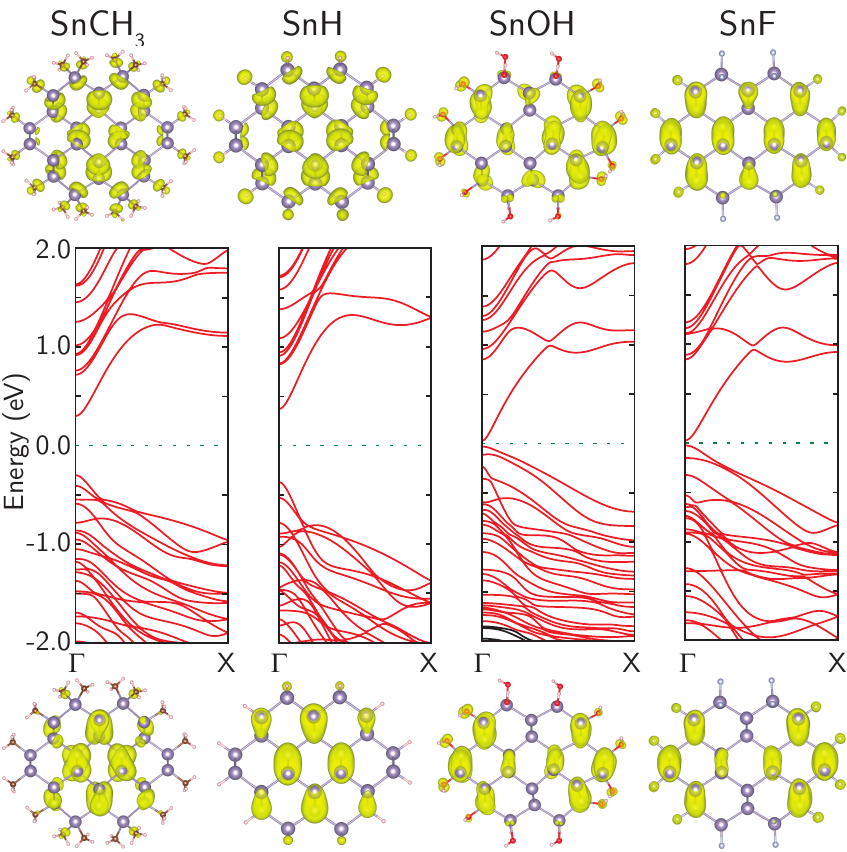}

\caption{Band structures of 1.5 nm $[110]$ $\alpha$-Sn NWs passivated with varying chemical species and 3D charge densities associated with valence band maxima (bottom) and conduction band minima (top). All charge densities taken at the same isosurface level. Zero of energy is set to the Fermi level.}\label{fig:bands2}

\end{figure}

\Cref{fig:bands2} shows the band structure of 1.5 nm $[110]$-oriented NWs for all surface terminations considered. Surface-induced variations in the potential confining electrons within the NW cores are observed to interact with different bands to varying degrees with larger shifts occurring in hydroxyl- and fluorine-passivated structures. Charge densities associated with the bottom of the conduction band and the top of the valence band of each structure are portrayed at the top and bottom of \cref{fig:bands2}, respectively. Shown charge distributions are similar for both structures passivated with terminations with lower electronegativity values, while charge distributions for structures with more electronegative passivants are similar amongst them. Inspection of charge distributions corresponding to the first few bands away from the gap reveals similarities across terminations for bands with different indices; this indicates modification of the confining potential induced by surface terminations can have an impact on band ordering --\emph{e.g.} the charge distribution of the valence band edge in hydroxyl- and fluorine-passivated structures matches that of the second topmost valence band in methyl- and hydrogen-passivated NWs--. In particular, none of the structures were found to present states near the Fermi level with charge densities localized at the surface indicating proper passivation of all Sn bonds.

\begin{table}[t]
\caption{\label{tab:GW}Band gap values for 1.5 nm $\alpha$-Sn NW structures as computed with DFT using pseudo-atomic orbitals (PAO), plane waves (PW), and PW calculations including first-order perturbative GW quasi-particle corrections.}

\begin{ruledtabular}
\begin{tabular}{l *{3}{c}}
& DFT-PAO & DFT-PW & GW \\
& \multicolumn{3}{c}{Band gap energy $\rm{\epsilon_G}$ (eV)} \\
\hline \\
$[ 100 ]$ &  &  & \\
\hspace{15pt} $\rm{H}$ & 1.61 & 1.62 & 3.54\\
$[ 110 ]$ &  \\
\hspace{15pt} $\rm{CH_{3}}$ & 0.54 & 0.53 & 1.77\\
\hspace{15pt} $\rm{H}$ &0.74 & 0.72 & 2.05\\
\hspace{15pt} $\rm{OH}$ & 0.02 & 0.02 & 0.25\\
\hspace{15pt} $\rm{F}$ & 0.03 & 0.01 & 0.70 \\

$[ 111 ]$ &  \\
\hspace{15pt} $\rm{H}$ &  1.11 & 1.12 & 2.85\\
\end{tabular}
\end{ruledtabular}
\end{table}

In order to estimate the order of magnitude of band gaps with increased accuracy, the DFT-level description is improved by means of first-order perturbative $GW$ quasi-particle value corrections. Corrections computed across the Brillouin zone for the conduction and valence band edges show a dependence with k-points in which the direct character exhibited by most of the structures in the study in DFT is maintained. This k-dependence indicates an explicit $GW$ calculation is required in order to compute the full quasi-particle band structure; in this study we have restricted our calculations to band gap magnitudes. \Cref{tab:GW} shows computed values of band gaps estimated using DFT with pseudo-atomic orbitals (PAOs) and plane-waves (PW) basis sets, and first order perturbative $GW$ corrected values for hydrogen-passivated SnNWs of all orientations considered, and for all other surface terminations for the case of $[ 110 ]$-oriented NWs, which DFT predicts to exhibit the lowest band gap values. The band gap values predicted from approximate DFT found with both basis sets are found to be in good agreement whereas $GW$ corrections significantly increase the magnitude of the energy band gaps. Even though non-selfconsistent $GW$ quasiparticle corrections are expected to underestimate energy band gaps values and depend on the starting DFT conditions, they have been shown to significantly improve upon DFT band gap descriptions.\cite{Schilfgaarde2006,Schilfgaarde2006b}

Trends observed in DFT across different surface passivants for $[ 110 ]$-oriented structures values are slightly modified after $GW$ corrections. While structures terminated with methyl and hydrogen continue to exhibit larger band gaps than those passivated with more electronegative terminations, hydroxyl-passivated SnNWs are predicted to exhibit a smaller energy band gap than the fluorine-passivated structure. Trends observed in DFT across crystallographic orientations for hydrogen-passivated SnNWs are maintained after $GW$ corrections with band gaps decreasing along the sequence $[ 100 ]- [ 111 ] - [ 110 ]$. Results indicate that $\alpha$-Sn NWs' band gap values can be greatly tuned with 1.5 nm structures exhibiting values ranging from 0.25 eV up to 3.54 eV, and variations greater than 1.5 eV achievable on single-crystalline structures by means of chemical modification. Intermediate band gap values are expected to be possible by combining more than one surface passivant in appropriate proportions, allowing a certain degree of \emph{band gap engineering} to be achieved through surface chemistry.\cite{Leu2006}

\subsection{Charge and potential}

\begin{figure}

\includegraphics[scale=1.0]{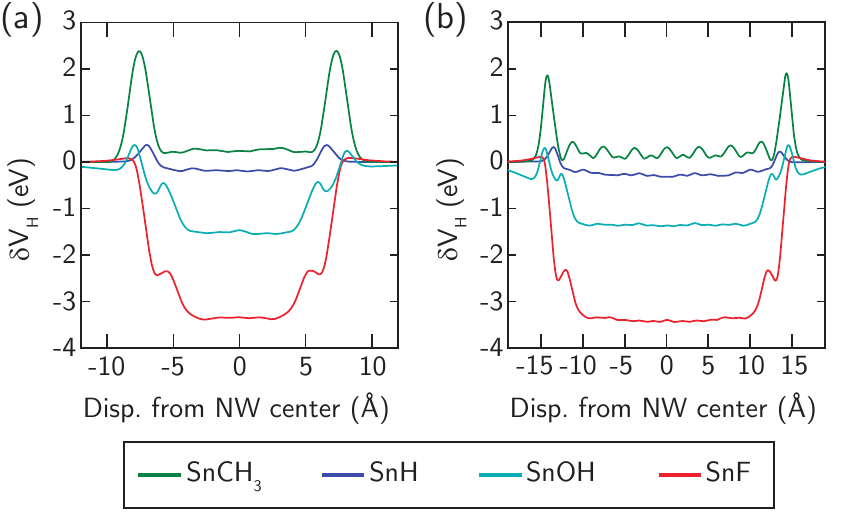}

\caption{Hartree difference potential profile across (a) 1.5 nm and (b) 3 nm NW cross sections. The depth of the induced well-like profile is found to increase with passivant electronegativity. Zero of energy has been fixed to potential values in regions far from the structures.}\label{fig:potentials}

\end{figure}

\begin{figure} 
\includegraphics[scale=1.0]{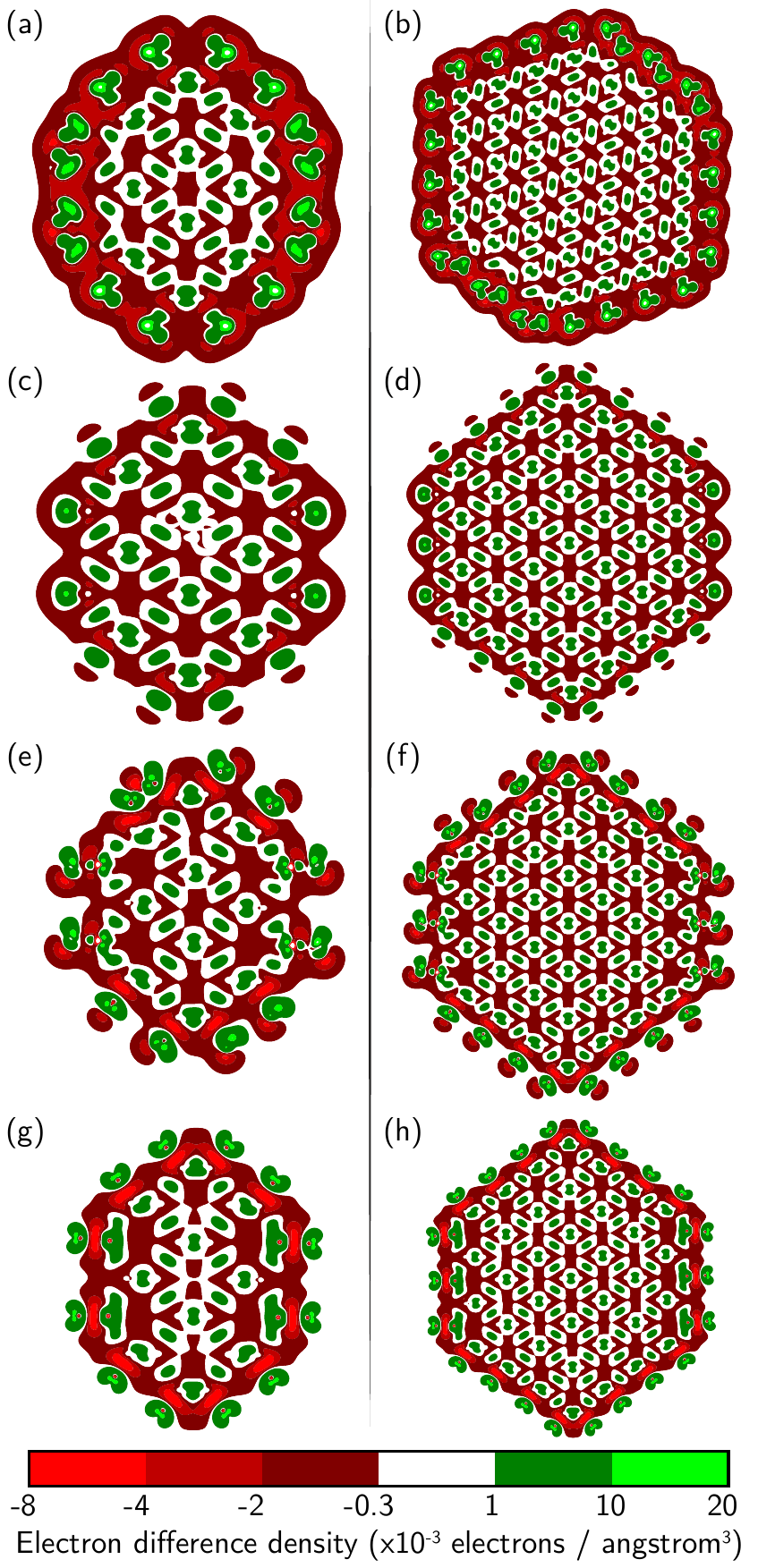}

\caption{Electron difference density averaged along NW axis for 1.5 nm and 3 nm structures oriented along $[ 110 [$ and passivated with (a)-(b) methyl groups, (c)-(d) hydrogen atoms, (e)-(f) hydroxyl groups, and (g)-(h) fluorine atoms.}\label{fig:charge}

\end{figure}

The modulation of the SnNW electronic properties achieved by means of chemical passivation presented in the previous subsection is analyzed in terms of the potential and charge distributions across the NWs. To support this discussion, the electrostatics in the 1.5 nm and 3 nm $[ 110 ]$ NWs is considered by examining the electron difference density $\delta n (\vec{r})$ defined as the difference in charge density between self-consistent DFT and the superposition of neutral atomic densities, and the Hartree difference potential $ \delta V_H ( \vec{r} ) $ defined as

\begin{equation}
\Delta \delta V_H [\delta n](\vec{r}) = - \frac{e^2}{4 \pi \epsilon_0} \delta n ( \vec{r} ).
\end{equation}

\Cref{fig:potentials} shows the Hartree difference potential across the cross-section of $[ 110 ]$ SnNWs with diameters of 1.5 nm and 3 nm NWs with different terminations averaged along a transverse axis. The profiles show potential wells profiles of varying depths depending on the electronegativity of the surface group with relatively shallow potential wells for the methyl- and hydrogen-terminated NWs relative to hydroxyl- and fluorine-terminated NWs; the latter surface terminations induce potential depths around an order of magnitude larger. The electrostatic potential flattens in regions near the center of the NWs with well depth comparable for NWs of different diameters and orientations but with the same terminating species.

Electron difference densities maps $\delta n (\vec{r})$ corresponding to these potentials are shown in \cref{fig:charge}. As Mulliken population analyses\cite{Mulliken1955} indicate, bonding to the Sn atoms withdraws charge from the NW onto the terminating groups according to their electronegativity with the Mulliken charge on hydrogen and methyl increasing by 0.10-0.15 electron and hydroxyl and fluorine's increasing by 0.5-0.6 electron. This is reflected in \cref{fig:charge} by regions of charge accumulation at the surface (green) corresponding to the atomic positions of the surface groups. Values around the Sn core indicate an overall deficit of electrons (red) with accumulation zones around bonds between atoms and especially near surface groups. In \cref{fig:charge}(e)-(h) green areas indicating accumulation of electrons can be seen near regions of strong charge deficit at the surface and towards the center of the NWs (i.e. \emph{inside} the Sn core); these correspond to the situation where the surface groups bond to surface Sn atoms parallel to the NW axis as depicted in \cref{fig:struct}(c)-(d) for hydroxyl-passivated $[ 110 ]$-oriented NW structures.

When electron withdrawal occurs at the surface of these systems, the Sn atoms' electrostatic screening results in a well-like potential profile, with the potential depth correlating to the amount of charge transfer taking place at the surface. A parallel may be drawn between these systems and the classical electrostatics problem of a charge density distributed along the surface of an infinitely long cylinder by treating electronic charge accumulation and the induced deficit in the NW core as two concentric cylindrical charge distributions. The charge distributions give rise to a potential difference between regions outside and inside both cylinders which can be expressed as

\begin{equation}
\Delta \varphi= \frac{n_s}{2 \pi \epsilon_0}\ln{\frac{r_o}{r_i}}
\end{equation}

where $n_s$ is the surface charge density, $r_o$ is the radius of the outer cylinder, $r_i$ the radius of the inner cylinder, and $\epsilon_0$ the permittivity of free space. By analogy, more electronegative passivating groups induce a larger surface charge density thus increasing the potential difference between regions outside and inside of the NW, which results in a net lowering of electronic states' energy and a concomitant increase in electron affinity. As discussed previously, $\Delta \varphi$ remains approximately constant across NWs with the same surface termination which points to induced surface charge densities which do not depend on the properties of the different surface facets exposed in NWs with different crystallographic orientations, but rather on the passivant-to-Sn ratio which is approximately constant for all structures of a given diameter.

Given the charge distribution inside the NW cores and their small cross-sections --of the order of the diameter of a benzene ring for the smaller NWs--, we may analyze the effect of modifying surface electronegativity by terminating groups on NW band gaps in terms of molecular electronics arguments first proposed by \citet{Aviram1974}. By establishing a parallel between molecules and NWs, and between substituent Hammett constants and passivant electronegativities\cite{Liu2004}, the observed trend of band gap reduction with increasing surface termination electronegativity as follows: the charge distribution induced by highly electronegative surface groups results in a NW core deficient in electrons, increasing the SnNW's electron affinity in the core thus lowering the energy of its conduction band --as with the lowest unoccupied molecular orbital (LUMO) in molecular rectifiers.

By extending the analogy with molecular electronics and exploiting the dependence of NW energy band gaps with surface terminations, the properties of a rectifying junction designed by abruptly switching surface groups from fluorine to hydrogen along the length of an infinitely long 1.5 nm diameter $[ 110 ]$-oriented NW has been studied.\cite{ASS2016B}.

\section{\label{Conclusion}Conclusion}

Electronic and structural properties of highly idealized crystalline $\alpha$-Sn NWs with diameters of approximately 1.5 nm and 3 nm oriented along $[ 100 ]$, $[ 110 ]$, and $[ 111 ]$ directions with surface bonding saturated by chemical groups of varying electronegativity are reported. Structurally, the NWs exhibit a remarkably weak dependence in total energy with respect to the lattice spacings along the NW axes. Surface terminations were not found to induce significant surface reconstructions or modify bond lengths in directions perpendicular to the NWs' axes, although bonding schemes were found to vary somewhat with different terminating species. The calculations show a high degree of control over the NWs' band gap and electron affinity can be achieved by varying the degree of charge transfer between NW core and surface groups by selecting the electronegativity of the latter. Charge transfer at the surface results in well-like potential profiles which increase in depth with increasing electronegativity of the surface groups lowering the energy of electronic states inside the NW core, increasing the system's electron affinity and reducing band gap values in most cases. Quasiparticle corrections within a $GW$ framework were performed to explore the range of band gap variations showing that sub-5-nm Sn structures can range from remaining semimetallic to widegap semiconductors via surface chemical modification. Surface groups allow counteracting quantum confinement effects and provide further control of electronic properties, allowing significant variations of energy band gap values along the length of structurally homogeneous SnNW cores via variation of surface coverages. Given the suitability of NW systems' electron transport properties for the design of nanoelectronics components\cite{Appenzeller2008} and the broad range of band gaps found for SnNWs we expect semimetallic systems in the sub-5-nm size range to be strong candidates for the design of next-generation charge-based components such as rectifiers and field-effect transistors.

\begin{acknowledgments}
This work was funded by Science Foundation Ireland through a Principal Investigator award Grant No. 13/IA/1956. ASS was supported by an Irish Research Council graduate fellowship. We acknowledge additional support from QuantumWise. Atomistic visualizations were rendered using VESTA \cite{VESTA}.
\end{acknowledgments}

\end{document}